\documentclass{aa}
\usepackage{txfonts}
\usepackage[dvipsnames]{xcolor}
\usepackage{graphicx}
\usepackage{placeins}
\usepackage{lscape}
\usepackage{ulem}
\usepackage{multirow}
\usepackage{hyperref}
\usepackage{tablefootnote}
\hypersetup{
    colorlinks=true,
    linkcolor=blue,
    citecolor=blue,
    filecolor=magenta,      
    urlcolor=blue,
    }

\def\say#1{`#1'}

\newcommand{\feh}{\mbox{\rm [{\rm Fe}/{\rm H}]}}

\newcommand{\Msun}{\mbox{$\mathrm{M}_{\odot}$}}

\newcommand{\ali}{\mbox{$\mathrm{A(Li)}$}}

\defcitealias{2025A&A...703A.204N}{Paper I}
\newcommand{\paperone}{\citetalias{2025A&A...703A.204N}\xspace}

\begin{document}

\title{\textit{Gaia}-Sausage-Enceladus: Lithium evolution from early red-giant-branch and main-sequence stars}
\titlerunning{\textit{Gaia}-Sausage-Enceladus: Lithium evolution from early red-giant-branch and main-sequence stars}

\subtitle{}

   \author{C. T. Nguyen
        \inst{1,2}
        \corrauth{chi.nguyen@inaf.it}
        \and
        G. Cescutti   
        \inst{1,2}
        \email{gabriele.cescutti@inaf.it}
        \and
        M. M. Bennedik
        \inst{3}
        \email{bennedik@astro.uni-tuebingen.de}
        \and 
        P. Molaro
        \inst{2}
        \email{paolo.molaro@inaf.it}
        \and 
        L. Magrini
        \inst{4}
        \email{laura.magrini@inaf.it}
        \and
        A. J. Korn         
        \inst{5}
        \email{andreas.korn@physics.uu.se}
          }

        \institute{
        Department of Physics, University of Trieste, Piazzale Europa, 1, Trieste, Italy
        \and
        INAF-Osservatorio Astronomico di Trieste, Via Giambattista Tiepolo, 11, Trieste, Italy
        \and
        Institut für Astronomie und Astrophysik, Eberhard Karls Universität Tübingen, Sand 1, 72076 Tübingen, Germany
        \and
        INAF, Osservatorio Astrofisico di Arcetri, Largo E. Fermi 5, 50125 Firenze, Italy
        \and
        Department of Physics and Astronomy, Uppsala University, Box 516, SE-75120 Uppsala, Sweden
      }

\authorrunning{Nguyen et al.}

   \date{}

\abstract{
The combination of data from the \textit{Gaia} satellite and large ground-based spectroscopic surveys recently lead to a milestone understanding of our Galaxy's formation history, marked by the identification of stellar remnants of the accreted \textit{Gaia}-Sausage-Enceladus (GSE) dwarf galaxy. 
Lithium (Li) remains one of the most difficult elements to explain because of its complex behaviour over evolutionary timescales: both the Spite plateau observed in metal-poor main-sequence (MS) stars and the recently discovered Li plateau of early red-giant-branch (eRGB) stars in the Milky Way challenge current galactic chemical evolution models. 
In this article, we investigate the viability of these Li-plateau features in the GSE galaxy, using public data from current big surveys: GALAH, \textit{Gaia}-ESO, and the collective SAGA database. We present a chemical evolution model of Li for GSE and find agreement with the observed data. 
We find the signature of Li plateau at low metallicities in both eRGB and MS stars. At higher metallicities, we see candidates of the Li-enriched stars that have their main contribution from nova explosions. These results reinforce the universality of the Spite plateau, and indicate that the eRGB Li plateau might also be a universal feature across different galactic systems. 
A hint of low nova Li yield in GSE is suggested by our eRGB sample from GALAH. 
However, the lack of stars at high metallicities, possibly caused by the merger event, prevents a precise study of nova contributions, and we expect that upcoming data will enable a more comprehensive analysis.}

   
\maketitle
\nolinenumbers
   
\section{Introduction}
Mergers with nearby galaxies have affected the structure and formation of the Milky Way (MW), which is formed via two main processes: in-situ formation and accretion. While in-situ stars were formed within the main body of the MW, accreted stars originated from nearby galaxies and were incorporated into it over time \citep[][]{2016ApJ...821....5D}. 
The most prominent event is the major merger of the \textit{Gaia}-Sausage-Enceladus (GSE) dwarf galaxy occurred roughly 10\,Gyr ago \citep{2018Natur.563...85H}. 
Thanks to the astrometric data from \textit{Gaia} \citep[][]{2016A&A...595A...1G} and the spectroscopic survey Apache Point Observatory Galactic Evolution Experiment \citep[APOGEE;][]{2017AJ....154...94M}, the identification of substructures in the MW stellar halo led to a profound discovery of the GSE stellar populations \citep[][]{2018Natur.563...85H,2018MNRAS.478..611B,2019MNRAS.488.1235M}. 
This merger event perturbed the formation and evolution of the MW \citep[][]{2026MNRAS.546ag111C}. 
Hence, it is crucial to understand the formation history of GSE progenitor. 
Historically, \citet{2010A&A...511L..10N} was the first to identify two distinct halo populations with high- and low-[$\alpha$/Fe] sequences, of which the lower [$\alpha$/Fe] sequence was argued to be an accreted population. This was later confirmed by the discovery of GSE. 
Since then, many studies have delved into the reconstruction of GSE star formation history using its stellar abundances \citep[e.g.][]{2019MNRAS.487L..47V,2021ApJ...923..172H}. 
Recently, by comparing the evolution of Mg, Fe, Ba and Eu with that of Sculptor and Fornax galaxies, \citet{2024A&A...691A.333E} suggested that GSE underwent a slow gas-enrichment first, then followed by a star formation phase and quenched by the merger. 
The metallicity range of GSE remnants is usually between ${-3\!\leq\!\feh\!\leq\!-0.6}$, with the metallicity distribution function (MDF) peaking at $\feh\!\approx\!-1.2$ \citep[][]{Feuillet_2021, 2022ApJ...935..109L}. 

Lithium (Li) is a fragile element, easily destroyed in the stellar interior at $\sim\!2.5\!\times\!10^6$\,K \citep[][]{1997ARA&A..35..557P}. Observations of warm and metal-poor main-sequence (MS) halo stars reveal a Li plateau, known as the Spite plateau \citep[][]{1982Natur.297..483S}, with an average abundance of $\ali\!\approx\!2.2$\,dex. In contrast, standard Big Bang Nucleosynthesis predictions \citep[][]{2014JCAP...10..050C}, assuming the cosmological parameters of \citet{2014A&A...571A..16P}, yield a higher primordial value, $\ali\!\approx\!2.7$\,dex, leading to the well-known \say{cosmological Li problem}. Stellar internal processes certainly play a role, as the stars we observe today have evolved over the past 10--12\,Gyr \citep[][]{2007ApJ...671..402K,2015MNRAS.452.3256F,2020MNRAS.497L..30G}. Meanwhile, stars at solar neighbourhood metallicity show a complex pattern of Li enrichment and depletion that may originate from many sources, e.g. cosmic rays \citep[][]{2012A&A...542A..67P}, nova explosions \citep[][]{2019MNRAS.482.4372C}, stellar mixing processes \citep[][]{2024A&A...690A.245B} and intrinsic properties \citep[][]{2025A&A...699A.173D}. 
These features make Li one of the most interesting elements to study galactic chemical evolution \citep[see][for a thorough review]{2026arXiv260217470C}. 

Recent study by \citet[][hereafter \paperone]{2025A&A...703A.204N} implemented corrections from stellar evolution as a possible explanation to the cosmological Li problem, and for the first time used early red-giant-branch (eRGB) stars to study Galactic Li evolution. Accordingly, chemical evolution model traces the evolution of an element in the interstellar medium. For Li, which is easily destroyed during the evolution of stars, a correction due to stellar evolution is thus embedded into the model to reproduce the abundances observed in stars at different evolutionary phases. 

Moreover, literature has confirmed the universality of the Spite plateau observed in MS stars in different galactic systems \citep[e.g.][]{2010A&A...519L...3M,2012A&A...543A..28N,2020MNRAS.496.2902M, 2021MNRAS.507...43S}. In this article, we first investigate the presence of the eRGB Li plateau in the GSE, and second present a galactic chemical evolution model for Li of the GSE. We adopt the available data from the Galactic Archaeology with HERMES survey \citep[GALAH,][]{2015MNRAS.449.2604D}, complemented by data from the SAGA database \citep{Suda_2008} and the \textit{Gaia}-ESO survey \citep[][]{2012Msngr.147...25G}. 
Details of the selection method for GSE members are described in Sect.~\ref{sample_selection}. In Sect.~\ref{cem}, we present a chemical evolution model for GSE. The obtained results are shown in Sect.~\ref{results}, and we conclude this paper in Sect.~\ref{conclusion}.

\section{Sample selection and kinematics}\label{sample_selection}
To study the Li evolution of GSE, we use data from the three catalogues GALAH~DR3 \citep{2024MNRAS.528.5394W}, SAGA \citep{Suda_2008} and \textit{Gaia}-ESO \citep{Magrini_2021}. We select the GSE members using the kinematic selection criteria defined by \citet{Feuillet_2021}, i.e. an angular momentum perpendicular to the galactic plane of ${-500\!\leq\!L_\mathrm{z} \,\rm{(kpc\,km\,s^{-1})}\!\leq\!500}$, and a radial action of ${30\!\leq\!\!\sqrt{J_\mathrm{r}\,\rm{(kpc\,km\,s^{-1})}}\!\leq\!55}$ (the so-called $L_\mathrm{z}$-$J_\mathrm{r}$ method). 

For stars in the GALAH~DR3 catalogue, we adopt the heliocentric coordinates and space velocities from \citet{2021MNRAS.506..150B}. For stars in the SAGA and \textit{Gaia}-ESO catalogues, these values are not provided and have to be computed. First, we seek the astrometric parameters from the \textit{Gaia}~DR3 catalogue \citep{GaiaCollaboration2023} by query at CDS using the star identification included in these catalogues. For radial velocities, we adopt the values provided in these catalogues if available, and from \textit{Gaia}~DR3 otherwise. Then, we compute the coordinates and space velocities from the astrometric parameters with the transformations outlined in Sect.~4.1.7.1 of \citet{GaiaDR3_DocChap4}.

We compute the orbital parameters with the Python package \textsc{galpy} \citep{Bovy_2015}. To be consistent with \citet{Feuillet_2021}, we use the \textsc{MWPotential2014} package and Stäckel approximation \citep{Binney_2012}, adopting the same distance of the Sun to Galactic centre ${R_\sun\!=\!8.0\,{\rm kpc}}$  \citep[][]{Bovy_2012}, its vertical offset to the Galactic plane $Z_\sun\!=25\,{\rm pc}$ \citep{Juric_2008} as well as the space velocities ${(U, V, W)_\sun\!=(11.1, 12.24, 7.25)\,{\rm km/s}}$ \citep{Schoenrich_2010} with a local standard of rest $V_{\rm LSR}\!=\!220\,{\rm km/s}$ \citep{Bovy_2012}. 
Using the kinematic $L_{\rm z}$-$J_{\rm r}$ selection method, we obtain 1257 stars from the GALAH DR3 catalogue (with high data quality: \texttt{flag\_fe\_h}=0 and \texttt{flag\_sp}=0), 32 stars from SAGA and 8 stars from \textit{Gaia}-ESO that belong to the GSE system. 
The stellar parameters, Li abundance and kinematic properties of these stars are listed in Tables~\ref{stellar_para_ALi} and \ref{kinematic_info}.

\begin{figure}
    \centering
    \includegraphics[width=1\columnwidth]{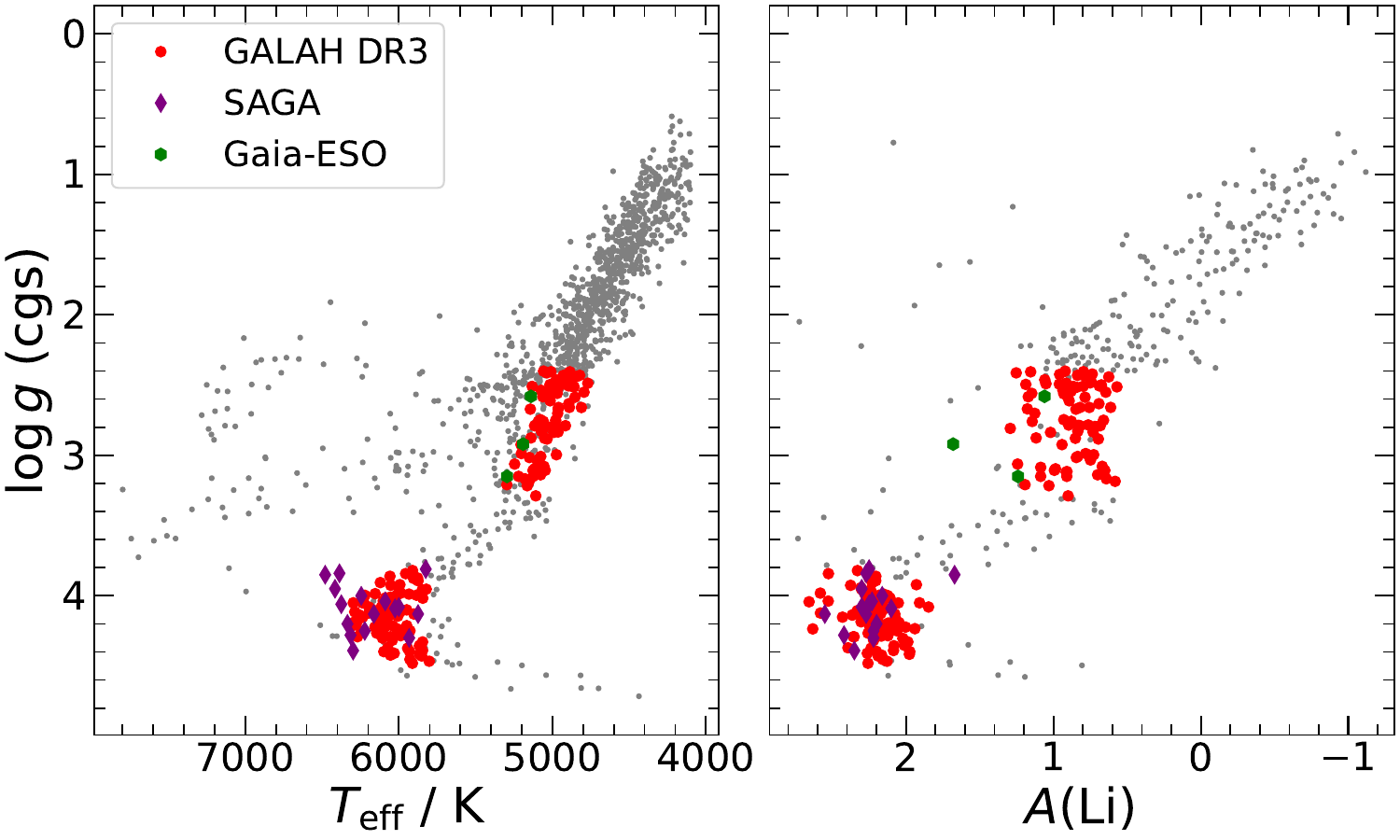}
    \caption{Samples of MS and eRGB stars of GSE from the GALAH, SAGA and \textit{Gaia}-ESO catalogues. Left panel: Kiel diagram. Right panel: Variation of A(Li) along with $\log g$. The full sample of GSE from the GALAH catalogue is shown in the background (grey dots). }
    \label{GSE_HRD}
\end{figure}

We further select the samples of MS and eRGB stars by applying conditions to $T_\mathrm{\!eff}$ and $\log g$. In particular, we select stars within ${5800\!\leq\! T_\mathrm{\!eff}\,({\rm K})\!\leq\!6900}$ and ${3.8\!\leq\!\log g\,({\rm cm\,s^{-2}})\!\leq\! 4.5}$ for our MS sample. For the eRGB stars, we define a synthetic line along the RGB phase within a narrow range of ${4800\!\leq\! T_\mathrm{\!eff}\!\leq\!5200}$ and ${2.4\!\leq\!\log g\!\leq\!3.3}$. Then, we select stars within the range of ${T_\mathrm{\!eff}^\mathrm{synthetic}\!\!\pm 200}$\,K for our eRGB sample. These choices of $T_{\rm eff}$  and $\log g$ are to avoid stars undergoing Li depletion due to convective driven and thermohaline mixing. The Kiel and $\ali$ vs. $\log g$ diagrams of stars in the two samples are shown in Fig.~\ref{GSE_HRD}. 

\section{Chemical evolution model}\label{cem}
In this article, we use the concept introduced in our previous work (\paperone), which takes into account corrections to surface Li from stellar evolution, to study the Li evolution of GSE,
\begin{align}\label{eq1}
    \mathrm{A(Li)} = \mathrm{A(Li)_{CEM}} + \Delta\mathrm{A(Li)_{SM}}.
\end{align}

The second term in Eq.~\ref{eq1}, $\Delta\mathrm{A(Li)_{SM}}$, is the correction due to stellar evolution. Similar to our previous work (for studying the Li evolution of the MW's thin disc), we adopt the grids of stellar corrections for surface Li from \citet{2025A&A...696A.136N}. The correction term is ultimately dependent on metallicity, since we adopt the empirical mass-metallicity relations in \paperone. More details on deriving $\Delta\mathrm{A(Li)_{SM}}$ can be found in appendix~\ref{appendix_DALiSM}.

\begin{table}[!tbp]
\caption{Computed values of the six free parameters and their best constrained values.
}
\label{free_parameters}
\centering
\begin{tabular}{l | c c}
\hline\hline
\multirow{2}{5em}{parameter} & range & \multirow{2}{6em}{best-constraint} \\
 & $([min, max, step])$ &  \\
\hline
$M_\mathrm{GSE}$ ($\Msun\,\mathrm{pc^{-2}}$) & $[1,\,3,\,0.5]$ & $2.38$ \\
$\tau$ ($\mathrm{Myr}$) & $[500,\,1000,\,50]$ & $729$ \\
$\sigma$ ($\mathrm{Myr}$) & $[1000,\,2000,\,100]$ & $1569$ \\
$\nu_\mathrm{SFR}$ ($\mathrm{Gyr^{-1}}$)& $[0.3,\,1.3,\,0.1]$ & $1.01$ \\
$\nu_\mathrm{wind}$ & $[3,\,9,\,1]$ & $6.83$ \\
$T_\mathrm{wind}$ ($\mathrm{Myr}$) & $[2100,\,3100,\,100]$ & $2900$ \\
\hline
\end{tabular}
\end{table}

The first term, $\mathrm{A(Li)_{CEM}}$, is the prediction from chemical evolution model. 
In this article, we adopt an infall model and assume a Gaussian infall rate to describe the accreted gas to form the GSE galaxy \citep[][]{2013A&A...553A..51C}. The infall rate is given by
\begin{align}
    \dot{G}_{\rm inf}(t) = M_\mathrm{GSE} \frac{e^{-(t-\tau)^2/2\sigma^2}}{\sigma\sqrt{2\pi}}.
\end{align}
Here, $M_\mathrm{GSE}$ is the total mass density of gas accreted into GSE, $\tau$ is the central peak of the infall distribution, and $\sigma$ is the standard deviation of the infall law. The star formation rate (SFR) follows the Schmidt's law \citep{1959ApJ...129..243S}, which is proportional to the surface gas density \citep[see also][]{1999A&A...350..827P}, ${\psi(t) = \nu_{\rm SFR}\Sigma(t)^k}$, 
if the evolutionary time of the GSE has not yet reached a threshold value, parametrized as $T_\mathrm{SFR}$. This value is set to the time when the GSE stops forming stars because of the interaction with the MW. As the evolutionary time goes beyond $T_\mathrm{SFR}$, the SFR of the GSE is set to zero. In the expression above, $\nu_\mathrm{SFR}$ is the efficiency coefficient of the SFR, $\Sigma(t)$ is the total surface mass density of gas in GSE, and the exponential coefficient $k=1.5$ following \citet{1998ApJ...498..541K}. Due to the interaction with the MW, an outflow wind operates for a period of time. In our model, the outflow wind rate is assumed to be proportional to the SFR, $W(t) = \nu_\mathrm{wind}\psi(t)$,  
with $\nu_\mathrm{wind}$ as the wind efficiency coefficient. The time when the wind starts is characterised by $T_\mathrm{wind}$, and before this time the outflow wind rate is set to zero. 

\begin{figure}
    \centering
    \includegraphics[width=1\columnwidth]{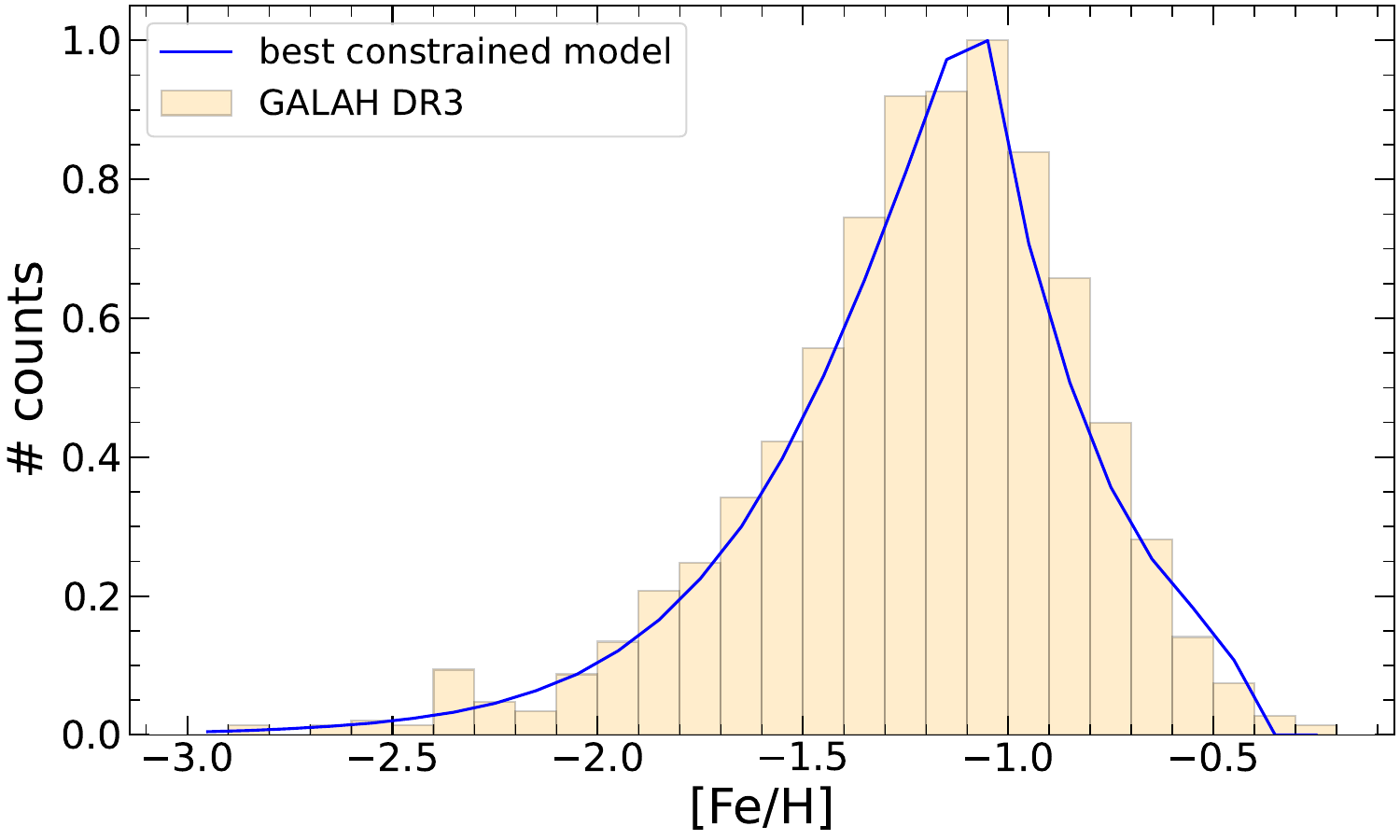}
    \caption{MDF of GSE star members. Data are taken from GALAH~DR3, binned in $0.1$\,dex of \feh. The displayed model is computed using the best-constrained values of the six parameters in Table~\ref{free_parameters}.}
    \label{GSE_MDF}
\end{figure}

In total, our chemical evolution model has seven free parameters: $M_\mathrm{GSE}$, $\tau$, $\sigma$, $\nu_\mathrm{SFR}$, $T_\mathrm{SFR}$, $\nu_\mathrm{wind}$ and $T_\mathrm{wind}$. The first calibration of these parameters was done in \citet{2020MmSAI..91..153C} using the APOGEE data. In this article, we recalibrate these parameters using GALAH~DR3 data. We stress that, in order to reduce the number of free parameters, we assume that GSE stopped forming stars at ${T_\mathrm{SFR}\sim 7}$\,Gyr ago. 
We then compute several models with given values of the other six parameters and rely on the metallicity distribution function (MDF) to search for the best-fit values. 
In particular, the parameter space of these six parameters are summarised in Table~\ref{free_parameters}. The normalised $\chi^2$ method is used to find the best match to the observed MDF. We collect the fits with ${\chi^2\leq 1}$ as our best-match models, and compute the mean values from the obtained results for each parameters. The best constrained values are listed in Table~\ref{free_parameters}. The MDF predicted by our model, computed with the best constrained values, is shown in Fig.~\ref{GSE_MDF} and superimposed on the observed MDF of GSE. 
Moreover, we assume the ISM's Li abundance is the cosmological value \citep[$\ali\!=\!2.69$\,dex;][]{2014JCAP...10..050C} in our calculations. 

Furthermore, Li enrichment in the solar vicinity is explained by the contribution of Li produced by novae. We adopt the prescription of \citet{2019MNRAS.482.4372C} to take into account their contribution. Here, we adopt the delay time between when the primary star became a white dwarf and the nova explosion takes place, $\tau_\mathrm{nova}$=$\,1$\,Gyr; we also adopt a mean nova Li yield, ${\rm ^{Li}Y_\mathrm{nova}\!=\!2.02\times\!10^{-5}\,\mathrm{M_\odot}}$, that is computed from observational data of Li production per nova event in the literature \citep[][]{2015Natur.518..381T,2016MNRAS.463L.117M,2016ApJ...818..191T,2018MNRAS.478.1601I,2020MNRAS.492.4975M,2021ApJ...916...44A,2022MNRAS.509.3258M,2023MNRAS.518.2614M,2025A&A...698A.291I}. We should clarify that an assumption that all novae produce the same amount of Li in all events is adopted in our model, together with a typical number of $10^4$ outburst events over their entire lifetime \citep[][]{1978ApJ...219..595F}. 
A computed model without any stellar Li depletion is shown by the pink line in Fig.~\ref{GSE_eRGB} for later comparisons.

\section{Results and discussion}\label{results}
\begin{figure}
    \centering
    \includegraphics[width=1\columnwidth]{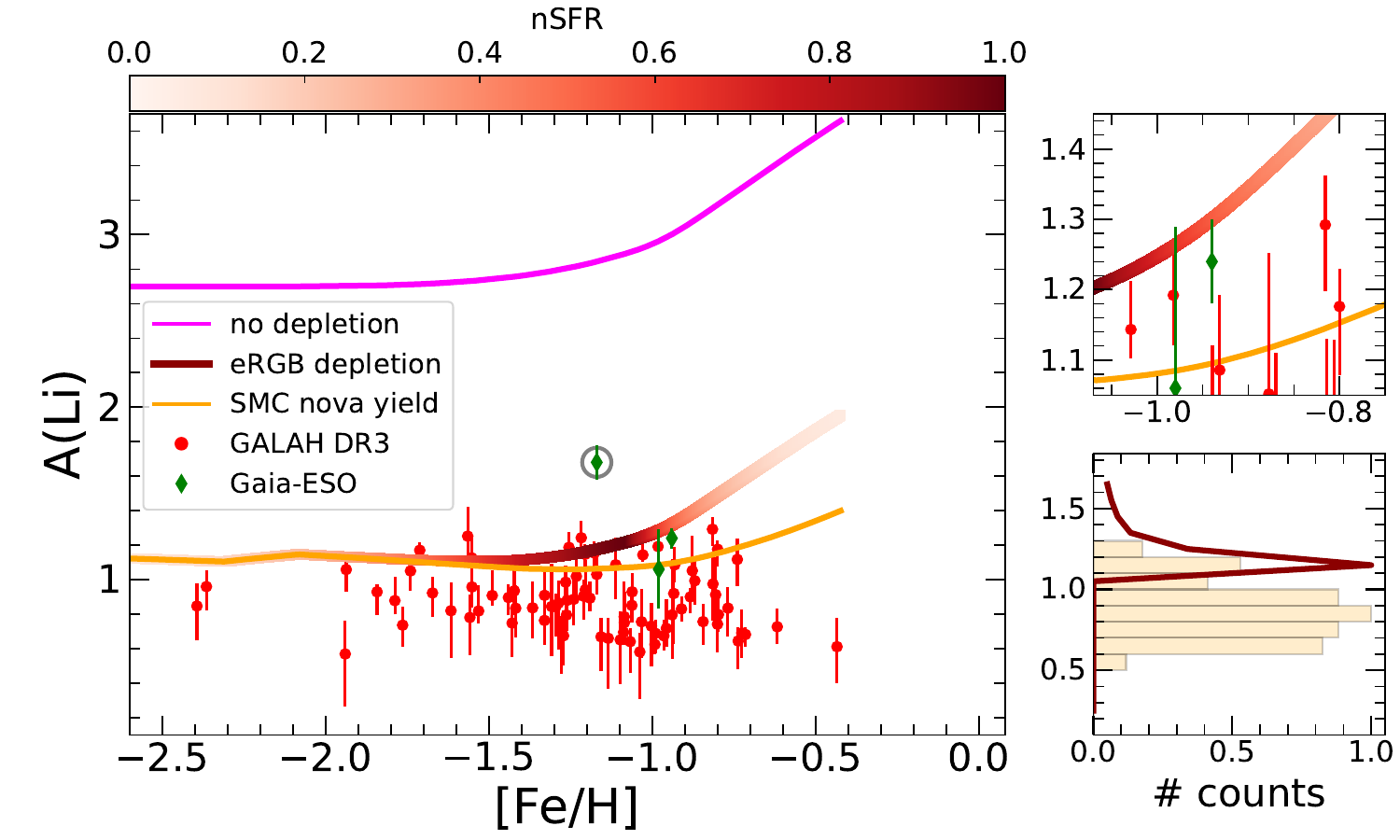}
    \caption{Left-panel: Li evolution of the eRGB stars. 
    Model with Li depletion and ${\rm ^{Li}Y_\mathrm{nova}\!=\!2.02\times\!10^{-5}\,\mathrm{M_\odot}}$ is colour-coded by the normalised SFR, while model with ${\rm ^{Li}Y_{nova}}\!=\!3.7\times 10^{-6}\,\Msun$ is shown by the orange line. 
    A possibly outlier star is circled in grey. 
    Upper-right panel: zoom-in to the region of $-1.05\!\leq\!\feh\!\leq\!-0.75$ of the left panel. Bottom-right panel: stellar distribution in Li abundance.}
    \label{GSE_eRGB}
\end{figure}

After constraining the free parameters using the MDF, we compute the chemical evolution model for Li assuming the original ${\ali\!=\!2.69\,\rm{dex}}$ and apply the correction term to take into account the Li depletion. Additional processes such as rotational mixing and angular momentum transport are expected to further enhance the Li depletion \citep[][]{2022NatAs...6..788E}, and intrinsic stellar properties may also contribute to the observed scatter \citep[][]{2025A&A...699A.173D}. As these effects are currently not included in the stellar correction term, our models preferentially trace the upper envelope of the evolutionary trend, explaining a small but systematic offset in the stellar distribution (Fig.~\ref{GSE_eRGB}, bottom panel).

In Fig.~\ref{GSE_eRGB} (left panel), we show the 82 eRGB stars of GSE selected from the GALAH~DR3 catalogue. It reveals a Li plateau at low metallicities, ${\feh\!\leq\!-1.2}$, with a mean value of $\ali=0.914$\,dex and a standard deviation of $0.15$\,dex, which is overlapping with the previous finding of \citet{2022A&A...661A.153M}.
The \citet{2022A&A...661A.153M} Li plateau was discovered among 58 field halo eRGB stars with high-solution spectra, and was reproduced by a Galactic thin disc chemical evolution model in \paperone. In this work, our model including the depletion correction up to the RGB bump is shown as the line coloured by the normalised SFR, indicating a Li-plateau feature at this metallicity range, 
with an average $\ali\!\approx\!1.1$\,dex. It is in agreement with the upper end of Li data from the GALAH sample. The predicted value is also well consistent with the result of \paperone and the observed plateau in \citet{2022A&A...661A.153M}. 
This result implies that the eRGB Li plateau may also be a universal feature across different galactic systems.

Novae are thought to play a dominant role in explaining the Li enrichment in the solar vicinity \citep[e.g.][]{2019MNRAS.482.4372C,2024A&A...689A.222K}. However, after the major merger event with the MW, GSE eventually stopped forming stars at a time possibly $T_\mathrm{SFR}\!\approx\!7$\,Gyr ago as we assumed in this work. This prevents the formation of younger stars that are potentially enriched in Li by the former nova explosions. As a result, the amount of GSE stars found in the solar neighbourhood are fewer towards higher metallicities, preventing us from having a better constraint on the nova Li yield. Regardless, we compute our model with an average Li yield of ${\rm ^{Li}Y_\mathrm{nova}\!=\!2.02\!\times\!10^{-5}\,\Msun}$ as constrained from the novae observations. 
Among the GALAH catalogue, at $\feh\!>\!-1.05$ region, we find that three most Li rich stars are aligning with prediction of our model (see Fig.~\ref{GSE_eRGB}, upper-right panel). Two of these stars, with $\feh\!\sim\!-1$, are located at the rising base of the enrichment evolution. The other star, $\feh\approx -0.8$, lies below our model prediction with an offset of $\sim 0.07$\,dex (including the observed uncertainty). 
We also show in Fig.~\ref{GSE_eRGB} three eRGB stars obtained from the \textit{Gaia}-ESO catalogue. Two of these stars show their abundances at the rising base of the enrichment evolution. 
The other star shows a relatively high Li abundance in comparison with other star members, marked by the grey circle. It has $\ali$=$1.68$\,dex, about $0.6$\,dex higher than the value reported by \citet{2022A&A...661A.153M}. We notice that its radial action, $\sqrt{J_\mathrm{R}}\!=\!30.08\,\mathrm{(kpc\,km\,s^{-1})^{1/2}}$, lies at the border of our selection criterion. This rises question about its membership. Moreover, the existence of Li-rich giants \citep[e.g.][]{2018A&A...617A...4S} further complicates the interpretation of its nature. We therefore consider it an outlier in this paper. 
The agreement between our model predictions and the Li abundances of these stars may help constrain the nova Li yield in GSE. However, they are mostly at the rising base of the enrichment evolution, preventing a firm conclusion on the adopted yield.

In addition, within the GALAH sample, a relatively flat trend can be depicted in our eRGB stars. This may suggest that GSE exhibits a lower nova Li yield. We include a model adopting the observed Li production per nova event of the Small Magellanic Cloud \citep[SMC;][]{2022MNRAS.510.5302I}, yielding ${\rm ^{Li}Y_{nova}}\!=\!3.7\times 10^{-6}\,\Msun$, in Fig.~\ref{GSE_eRGB} (orange line). The MDF of SMC peaks at $\feh\!\sim\!-1$ \citep[][]{2023A&A...671A.124M}, which is similar to the GSE metallicity. Our model with low nova Li yield tends to reproduce the flat tendency in eRGB stars, but is quite low to reproduce the most Li rich stars. 
We should stress that nova Li yield depends on stellar mass and metallicity, and that more work is needed to capture the full complexity of chemical enrichment from a population of novae, particularly for GSE, where stars at higher metallicities are lacking \citep[see][for more discussion]{2022MNRAS.509.1175K}.

\begin{figure}
    \centering
    \includegraphics[width=1\columnwidth]{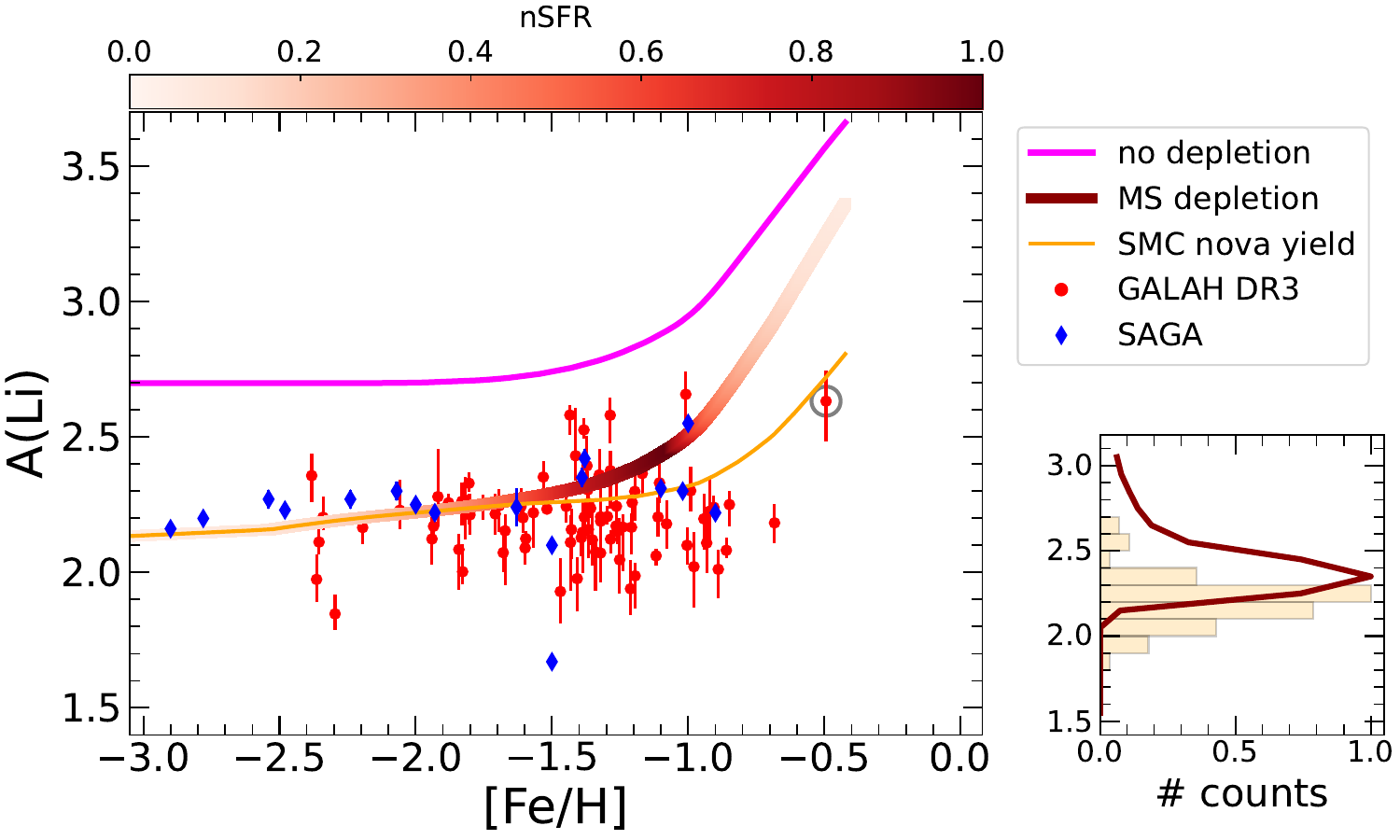}
    \caption{Same as Fig.~\ref{GSE_eRGB} but for the MS stars.}
    \label{GSE_MS}
\end{figure}

The Spite plateau found in the MW had been suggested as a common feature also in other galaxies \citep[e.g.][]{2010A&A...519L...3M}. Likewise, \citet{2020MNRAS.496.2902M} (hereafter, MCF20) drew a similar conclusion on GSE by adopting data from the GALAH~DR2 and SAGA catalogues, aligning with the conclusions of \citet{2012A&A...543A..28N}. However, MCF20 determined the GSE members by cross-matching with the former determination of \citet{2018Natur.563...85H}. In this regard, we recall that we use the new definition of \citet{Feuillet_2021} to determine the memberships of GSE (Sect.~\ref{sample_selection}). 
Due to differences in the definition, we find only 15 stars in our SAGA sample present in the MCF20 sample. 
Regardless of this limited overlap, we find agreement with MCF20 on the Spite plateau in GSE. 
In particular, at $\feh\!<\!-1.4$, our model shows an excellent fit to the observed data, especially of SAGA (Fig.~\ref{GSE_MS}, left panel), with a predicted $\ali\!\approx\!2.14-2.31$\,dex. Our data at this metallicity range from the GALAH sample indicates a mean Li abundance of $\ali\!=\!2.2$\,dex and a standard deviation of $0.14$\,dex. Furthermore, \citet{2021MNRAS.507...43S} applied the same GSE selection method to the GALAH~DR3 data \citep[][]{2021MNRAS.506..150B} and found that GSE indeed exhibits the same Li plateau ($\ali\!=\!2.35\pm 0.12$\,dex) as other accreted and in-situ MW stars. 
This result and the agreement with other studies once again reinforces the universality hypothesis of the Spite plateau across different galactic systems \citep[see also][]{2021MNRAS.505..200M}. 

Similarly to the case of eRGB stars, we could not fully study the contribution of novae at the enrichment evolution by using MS stars. We find only two stars at $\feh\! \sim\!-1$, each from one catalogue, located at the rising base of the enrichment branch. Nonetheless, our model, adopting the mean value of observed nova Li yields, predicts the abundances of these two stars very well. 
The most metal rich star from the GALAH sample in Fig.~\ref{GSE_MS} shows a Li abundance that can be reproduced by lowering the nova Li yield, as indicated by our model with a yield of the SMC. 
However, radial action of this star, $\sqrt{J_{\rm r}}\!=30.3\,\mathrm{(kpc\,km\,s^{-1})^{1/2}}$, raises questions about its membership. Therefore, we do not rely on this star in our current calibration. 
In this regard, we note that current and ongoing large surveys are rapidly improving both in data precision and sample sizes. We thus remain optimistic about more robust studies in the future.

Additionally, we should clarify that the GALAH DR4 catalogue \citep[][]{2025PASA...42...51B} recently releases a larger sample of stars. However, the catalogue has a problem in the flagging of [Fe/H], so we cannot select stars with a good determination of iron abundances. For this reason, we chose to adopt data from the GALAH DR3 catalogue in this work. A tentative test using GALAH DR4 data can be found in appendix \ref{DR4_redo}. In general, we see similar trends in the A(Li) versus [Fe/H] diagram between the two catalogues.

\section{Summary and conclusions}\label{conclusion}
The CEM of Li in a dwarf galaxy is expected to be different from that of the MW. 
In this context, the accreted GSE stars offer a unique opportunity to study Li evolution in a dwarf galaxy, since dwarf galaxies are often distant and their stellar Li abundances are difficult to measure. We adopt the available data from recent and current big surveys: GALAH~DR3, \textit{Gaia}-ESO, and the SAGA database; and use the $L_\mathrm{z}$-$J_\mathrm{r}$ method to determine GSE membership. We then present the Li evolution models for GSE and compare them with the observed data to study the Li plateaus in the MS and eRGB phases. The results reinforce the universality of the Spite plateau and suggest that the eRGB Li plateau may also be universal across different galactic systems. This opens a new category of observational targets for galactic Li evolution studies, since eRGB stars are significantly brighter and easier to observe at high resolution than near-MS stars. 
Our model also leads to a hint that GSE may exhibit a low nova Li yield, and that improvements in both the modelling of nova contributions and data sample would help to draw more robust conclusions on this matter. 
To conclude this article, we present in Tables~\ref{stellar_para_ALi} and \ref{kinematic_info} the stellar parameters and kinematic information of GSE stars determined in this paper, available at a Zenodo repository\footnotemark{}.
\footnotetext{\url{https://doi.org/10.5281/zenodo.19736686}}

\begin{acknowledgements}
We are grateful for the kind and constructive comments and suggestions from the referee, especially on the discussion of nova Li yield. This certainly strengthens our paper. 
CTN, GC, LM acknowledge the financial support under the
National Recovery and Resilience Plan (NRRP), Mission 4, Component 2,
Investment 1.1, Call for tender No. 104 published on 2.2.2022 by the Italian Ministry of University and Research (MUR), funded by the European Union – NextGenerationEU – Project ‘Cosmic POT’ (PI: L. Magrini) Grant Assignment Decree No. 2022X4TM3H by the Italian Ministry of Ministry of University and Research (MUR). 
CTN, GC, AK acknowledge funding from the European Union’s Horizon 2020 research and innovation programme under grant agreement No 101008324 (ChETEC-INFRA). 
CTN, GC acknowledge the support by INAF Mini grant 2024, ``GALoMS – Galactic Archaeology for Low Mass Stars'' (1.05.24.07.02). 
AJK acknowledges support by the Swedish National Space Agency (SNSA). 
The authors gratefully acknowledge the works behind the {\it Gaia} mission, ground-based spectroscopic surveys: GALAH, \textit{Gaia}-ESO, and the collective SAGA database, for providing the data used in this work. 
This work also makes use of the \texttt{TOPCAT} software \citep[][]{2005ASPC..347...29T}. 
CTN thanks Emanuele Spitoni, Alexandro Saro, Emma Dodd, Ella Xi Wang, Linda Lombardo and Umberto Maio for helpful discussions over the course of this work. 
\end{acknowledgements}

\bibliographystyle{aa}
\bibliography{references} 

\begin{appendix}
\nolinenumbers

\begin{table*}[h!]
\section{Table}\label{appendix_table}
Here we present the stellar parameters and Li abundances of 32 stars from the collective SAGA database, along with their references in Table~\ref{stellar_para_ALi}. Meanwhile, for stars from the GALAH \citep[][]{2024MNRAS.528.5394W} and Gaia-ESO \citep[][]{Magrini_2021} catalogues, we refer readers to the original catalogues for these properties. In Table~\ref{kinematic_info}, we summarise the kinematic and orbital properties of all GSE stars from three catalogues, determined in this article. These two tables can be found at the mentioned Zenodo repository\footnotemark[\value{footnote}].\\

\caption{Stellar parameters and Li abundances of 32 stars from the SAGA database that belong to the GSE. The last column reports references.}
\label{stellar_para_ALi}
\resizebox{1.0\textwidth}{!}{
\begin{tabular}{llccccccc}
\hline\hline
Gaia DR3 ID & Name & $\rm{T_{eff}}$ & log g & A(Li) & err & [Fe/H] & err & Ref. \\
 & & (K) & ($\rm{g\,cm^{-2}}$) & (dex) & (dex) & (dex) & (dex) & \\
\hline
4376174445988280576 & BD+02\_3375 & 6163 & 4.13 & 2.27 & 0.035 & -2.24 & --- & [1] \\
1776289248313154688 & BD+17\_4708 & 6025 & 4.09 & 2.1 & 0.04 & -1.5 & 0.079 & [2] \\
866863321051682176 & BD+24\_1676 & 6387 & 3.84 & 2.27 & 0.035 & -2.54 & --- & [1] \\
4005317278538492928 & BD+26\_2251 & 5875 & 4.13 & 2.55 & 0.05 & -1.0 & 0.079 & [2] \\
1458016709798909952 & BD+34\_2476 & 6416 & 3.95 & 2.3 & 0.035 & -2.07 & --- & [1] \\
761871677268717952 & BD+36\_2165 & 6315 & 4.28 & 2.42 & 0.035 & -1.38 & --- & [1] \\
5762455439477309440 & BD-03\_2525 & 5750 & 3.6 & 1.96 & 0.11 & -1.8 & 0.079 & [2] \\
3202470247468181632 & BD-06\_855 & 5378 & 4.43 & 0.98 & 0.04 & -0.69 & 0.05 & [2] \\
5510893810476230144 & CD-45\_3283 & 5672 & 4.88 & < 0.7 & --- & -0.83 & 0.05 & [2] \\
5551565291043498496 & CD-48\_2445 & 6222 & 4.25 & 2.22 & 0.01 & -1.93 & 0.05 & [3] \\
5486881507314450816 & CD-57\_1633 & 5933 & 4.3 & 2.22 & 0.05 & -0.9 & 0.05 & [2] \\
6573170751851975936 & CS22881-039 & 5950 & 2.1 & < 0.6 & --- & -2.75 & 0.19 & [4] \\
6663163201607650688 & CS22947-187 & 5200 & 1.5 & < 0.5 & --- & -2.61 & 0.09 & [5] \\
6863245112083068032 & CS22950-173 & 6335 & 4.2 & 2.199 & 0.09 & -2.78 & 0.121 & [6] \\
2676443097097288704 & CS22965-054 & 6245 & 4.0 & 2.161 & 0.09 & -2.9 & 0.136 & [6] \\
2601354871056014336 & CS29512-073 & 5600 & 3.4 & 1.93 & 0.1 & -2.14 & 0.14 & [5] \\
1001659495247619584 & G192-43 & 6298 & 4.39 & 2.35 & 0.035 & -1.39 & --- & [1] \\
1406910065012679680 & G202-65 & 6480 & 3.85 & 1.67 & 0.04 & -1.5 & 0.16 & [7] \\
2688113962054816128 & G26-12 & 6089 & 4.04 & 2.23 & 0.05 & -2.48 & 0.08 & [2] \\
2502689198705422848 & G75-31 & 6000 & 4.08 & 2.3 & 0.01 & -1.02 & 0.05 & [3] \\
3246885466349457536 & HD024289 & 5700 & 3.5 & 2.19 & 0.08 & -2.22 & 0.1 & [7] \\
615943806835727872 & HD084937 & 6375 & 4.06 & 2.25 & 0.028 & -2.0 & 0.079 & [2] \\
2453397508316944128 & HE0134-1519 & 5500 & 3.2 & 1.27 & 0.19 & -3.98 & 0.3 & [8] \\
4868057499705138560 & HE0440-3426 & 4800 & 1.6 & < 0.26 & --- & -2.19 & 0.2 & [9] \\
2988284007989400576 & J051727.4-134235 & 4961 & 1.93 & 0.91 & 0.1 & -2.14 & 0.01 & [10] \\
5602882389229755776 & J070520.3-334324 & 4757 & 1.27 & 0.81 & 0.1 & -2.24 & 0.02 & [10] \\
29184921251610880 & LAMOSTJ030209.33+135656.3 & 5206 & 2.3 & 2.34 & 0.12 & -1.74 & 0.3 & [11] \\
577295698241539840 & SDSSJ090733+024608 & 5934 & 3.71 & 2.23 & 0.1* & -3.42 & 0.09 & [12] \\
4715919175280799616 & CD-61\_282 & 5909 & 4.59 & 2.12 & 0.05 & -1.15 & 0.05 & [2] \\
4272653983123701120 & G21-22 & 5869 & 3.93 & 2.24 & 0.07 & -1.63 & --- & [13] \\
3510294882898890880 & HD111980 & 5850 & 3.94 & 2.31 & 0.1 & -1.1 & 0.08 & [14] \\
3374633977170787072 & HD250792 & 5568 & 4.4 & 1.5 & 0.035 & -1.01 & --- & [1] \\
\hline
\end{tabular}}
\tablebib{
[1]~\citet{2010A&A...515L...3M}; 
[2]~\citet{2009A&A...499..103S}; 
[3]~\citet{2006ApJ...644..229A}; 
[4]~\citet{2011A&A...527A..65H}; 
[5]~\citet{2012ApJ...751...14M}; 
[6]~\citet{2010A&A...522A..26S}; 
[7]~\citet{2007ApJ...667.1196B}; 
[8]~\citet{2014ApJ...787..162H}; 
[9]~\citet{2015ApJ...807..173H}; 
[10]~\citet{2018ApJ...868..110S}; 
[11]~\citet{2018ApJ...852L..31L}; 
[12]~\citet{2012A&A...542A..87B}; 
[13]~\citet{1997MNRAS.285..847B}; 
[14]~\citet{2009MNRAS.392..205T};
* private communication. 
}
\end{table*}

\begin{table*}[h!]
\caption{Kinematic and orbital parameters of GSE stars in this article. The last four columns are angular momentum in $z$ direction ($\rm{L_z}$), total orbit energy ($\rm{E_n}$), radial action ($\rm{J_r}$) and vertical action ($\rm{J_z}$), computed in Sect.~\ref{sample_selection}. }
\label{kinematic_info}
\centering
\resizebox{1.0\textwidth}{!}{
\begin{tabular}{lcccccccc}
\hline\hline
Name & RA & DEC & Plx & $\cdots$ & ${\rm L_z}$ &  ${\rm E_n}$ &  ${\rm J_r}$ &  ${\rm J_z}$ \\
  & (deg) & (deg) & (mas) & $\cdots$ & ($\rm{kpc\,km\,s^{-1}}$) & ($\rm{km^2\,s^{-2}}$) & ({$\rm{kpc\,km\,s^{-1}}$}) & ({$\rm{kpc\,km\,s^{-1}}$}) \\
\hline
SAGA BD+02\_3375 & 264.9383511 & 2.416891479 & 9.8386 & $\cdots$ & -63.47 & -3873.27 & 2839.32 & 33.59 \\
SAGA BD+17\_4708 & 332.8830976 & 18.09309066 & 7.6143 & $\cdots$ & -401.07 & -15886.19 & 1918.38 & 0.30 \\
$\vdots$ & $\vdots$ & $\vdots$ & $\vdots$ & $\vdots$ & $\vdots$ & $\vdots$ & $\vdots$ &$\vdots$ \\
GESO J01103865-5008389 & 17.66100446 & -50.14429859 & 0.3975 & $\cdots$ & -160.74 & -29247.54 & 1453.31 & 90.23 \\
GESO J04392439-4234170 & 69.85173290 & -42.57148078 & 0.283 & $\cdots$ & -293.27 & -13440.95 & 2121.02 & 144.28 \\
$\vdots$ & $\vdots$ & $\vdots$ & $\vdots$ & $\vdots$ & $\vdots$ & $\vdots$ & $\vdots$ &$\vdots$ \\
GALAH 131116000501201 & 51.44879759375034 & -68.64119851328851 & 1.6468 & $\cdots$ & -145.62 & -18877.52 & 1841.05 & 41.21 \\
GALAH 131120002501212 & 61.66282556896425 & -62.03528672946472 & 0.9815 & $\cdots$ & 366.32 & -24478.15 & 1480.94 & 27.07 \\
$\vdots$ & $\vdots$ & $\vdots$ & $\vdots$ & $\vdots$ & $\vdots$ & $\vdots$ & $\vdots$ &$\vdots$ \\
\hline
\end{tabular}
}
\end{table*}

\FloatBarrier 
\twocolumn

\section{The stellar correction term}
\label{appendix_DALiSM}
We here summarise the derivation of $\Delta\mathrm{A(Li)_{SM}}$ in Eq.~\ref{eq1}. The term is computed from grids of stellar tracks, providing the Li depletion from initial value, reflecting Li abundance of the ISM, to a given evolutionary phase (e.g. the MS and eRGB phases).

First of all, it should be recalled that the CEM calculation traces the evolution of a given element in the interstellar medium (ISM; or the stellar birth cloud) across galactic lifetime. Details of CEM for GSE can be found in Sect.~\ref{cem} of this paper. The Li abundance of ISM, whether cosmological ($\ali\!\approx\!2.7$\,dex) or Spite plateau ($\ali\!\approx\!2.2$\,dex) values, is a debatable subject still to date \citep[see also][]{2024A&A...690A..38M}. However, in this paper, we assume the Li abundance of ISM is the cosmological value and do not necessarily disown the other scenario, as we showed in \paperone.  

For deriving $\Delta\mathrm{A(Li)_{SM}}$ in the eRGB phase, we adopt the stellar grid of \citet{2025A&A...696A.136N} which includes two sets of tracks of metallicities, $\feh=-2.4$ and $-2.1$, and masses from $0.65$ to $0.94\,\Msun$. 
The predicted stellar-evolution correction is equal to the difference between the initial value in the beginning of the PMS phase and the value when the star reaches the Li-plateau value among RGB stars at a $\log g$ of $\sim\!2.8\,\rm{(cm\,s^{-2})}$. 
We should clarify that these tracks were computed with an initial $\ali =2.69$\,dex, consistent with the assumed $\ali$ of ISM in our CEM, and subsequently depleted during the PMS phase so that they reach the Spite plateau value at the MS phase. The stellar model prediction was then calibrated to Li data of the globular cluster NGC~6397. 
Additionally, if the ISM is assumed to have the Spite plateau value, there would be no depletion during the PMS phase as definition. 
In this case, the predicted stellar-evolution correction is equal to the difference between the value in the MS phase and the value in the RGB phase as defined above. 
Finally, due to the limited range of metallicity, an extrapolation scheme was applied in case the galactic evolutionary metallicity exceeds the limits of the stellar grid. 

A similar approach is applied to derive $\Delta\mathrm{A(Li)_{SM}}$ in the MS phase. In addition, we must emphasise that there are other mechanisms have been proposed to explain the cosmological Li problem by introducing different stellar extra-mixing prescriptions such as turbulent diffusion \citep[][]{2005ApJ...619..538R}, late mass accretion during the PMS phase \citep[][]{2015MNRAS.452.3256F} or rotation and magnetic fields \citep[][]{2026arXiv260222516Y}. Access to these stellar libraries would be valuable for deriving the stellar correction term in our CEM; however, we are at present limited in this regard.

\section{Analysis using GALAH DR4 data}\label{DR4_redo}
\begin{figure}
    \centering
    \includegraphics[width=1\columnwidth]{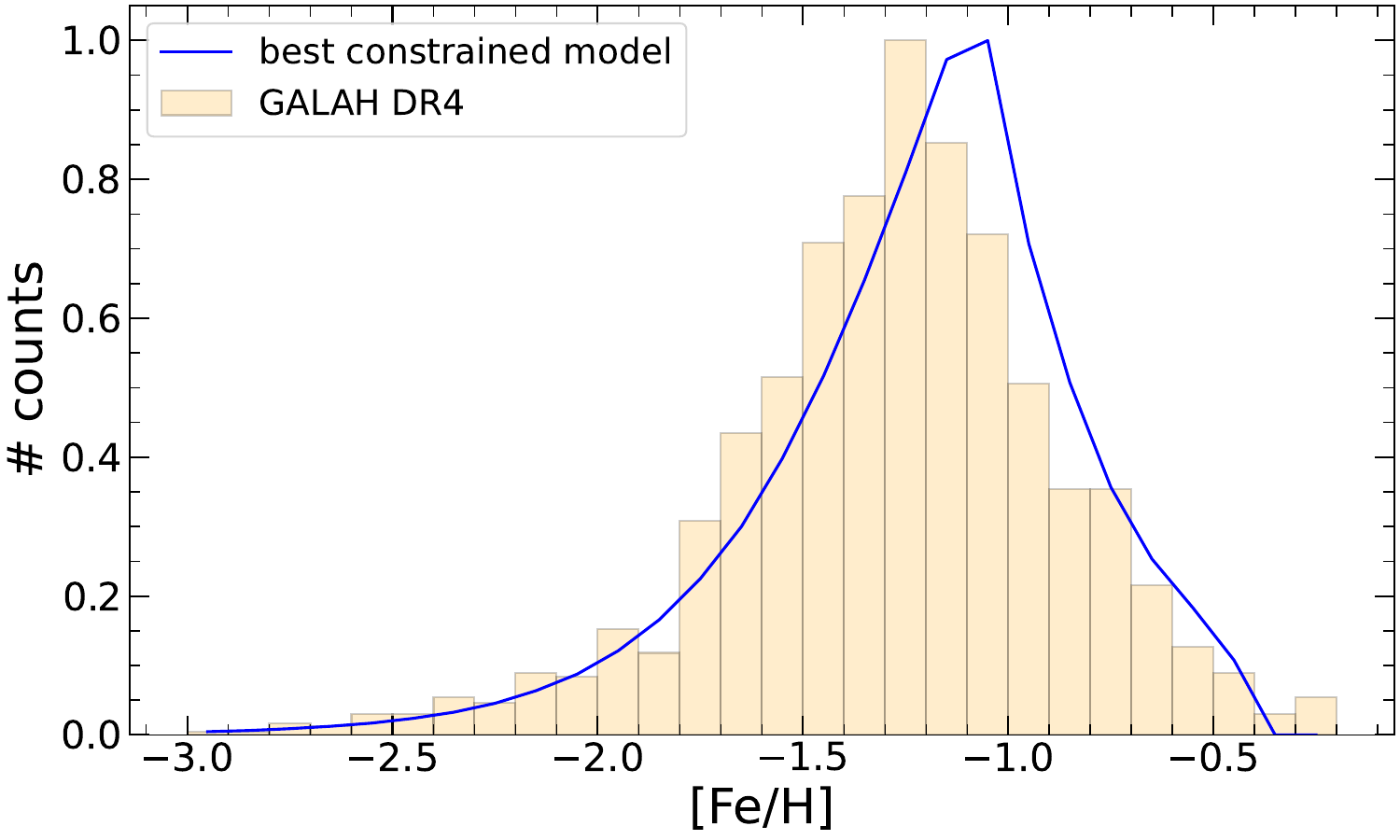}
    \caption{MDF of GSE by using the GALAH DR4 data.}
    \label{DR4_GSE_MDF}
\end{figure}

\begin{figure}
    \centering
    \includegraphics[width=1\columnwidth]{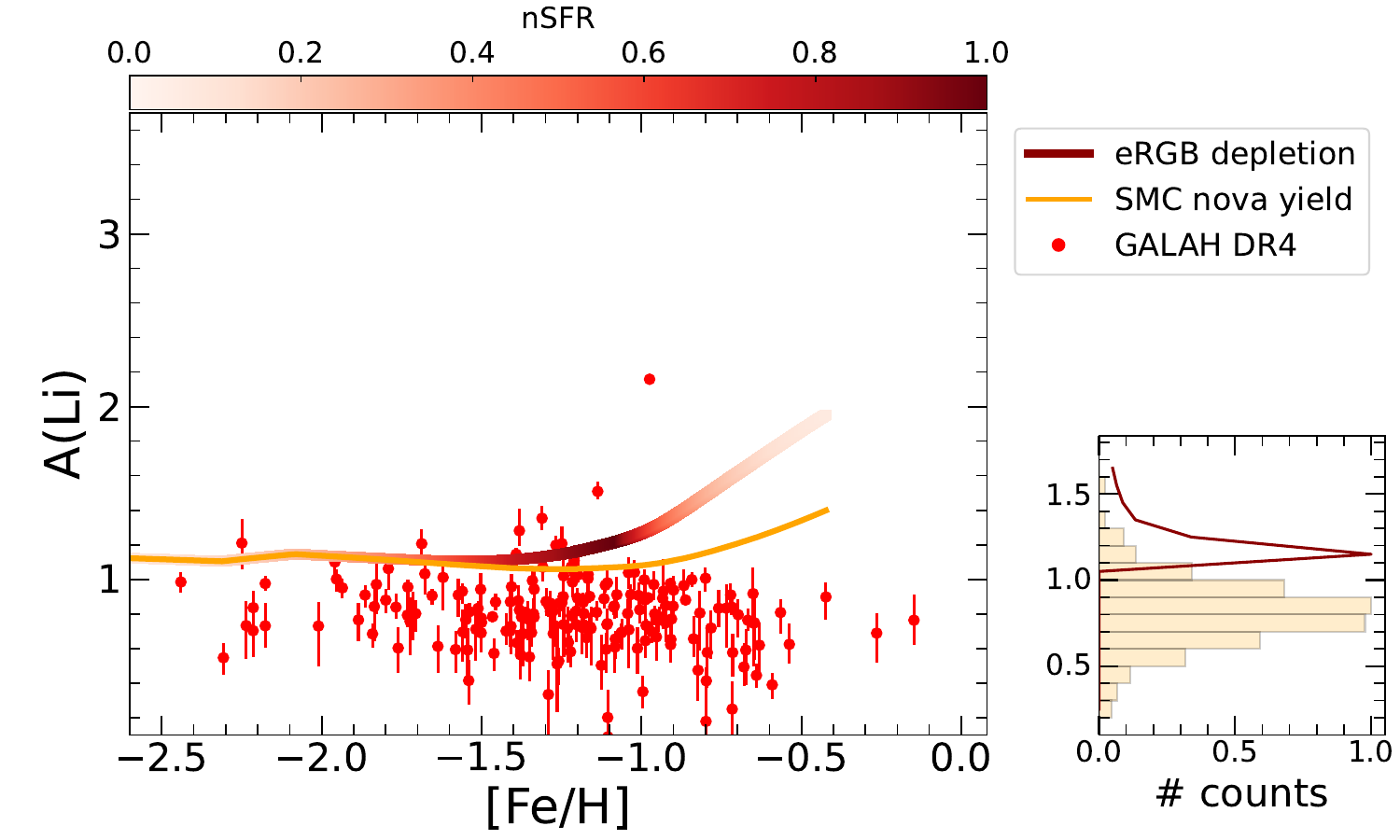}
    \caption{Li evolution in eRGB stars by using the GALAH DR4 data.}
    \label{DR4_GSE_eRGB}
\end{figure}

\begin{figure}
    \centering
    \includegraphics[width=1\columnwidth]{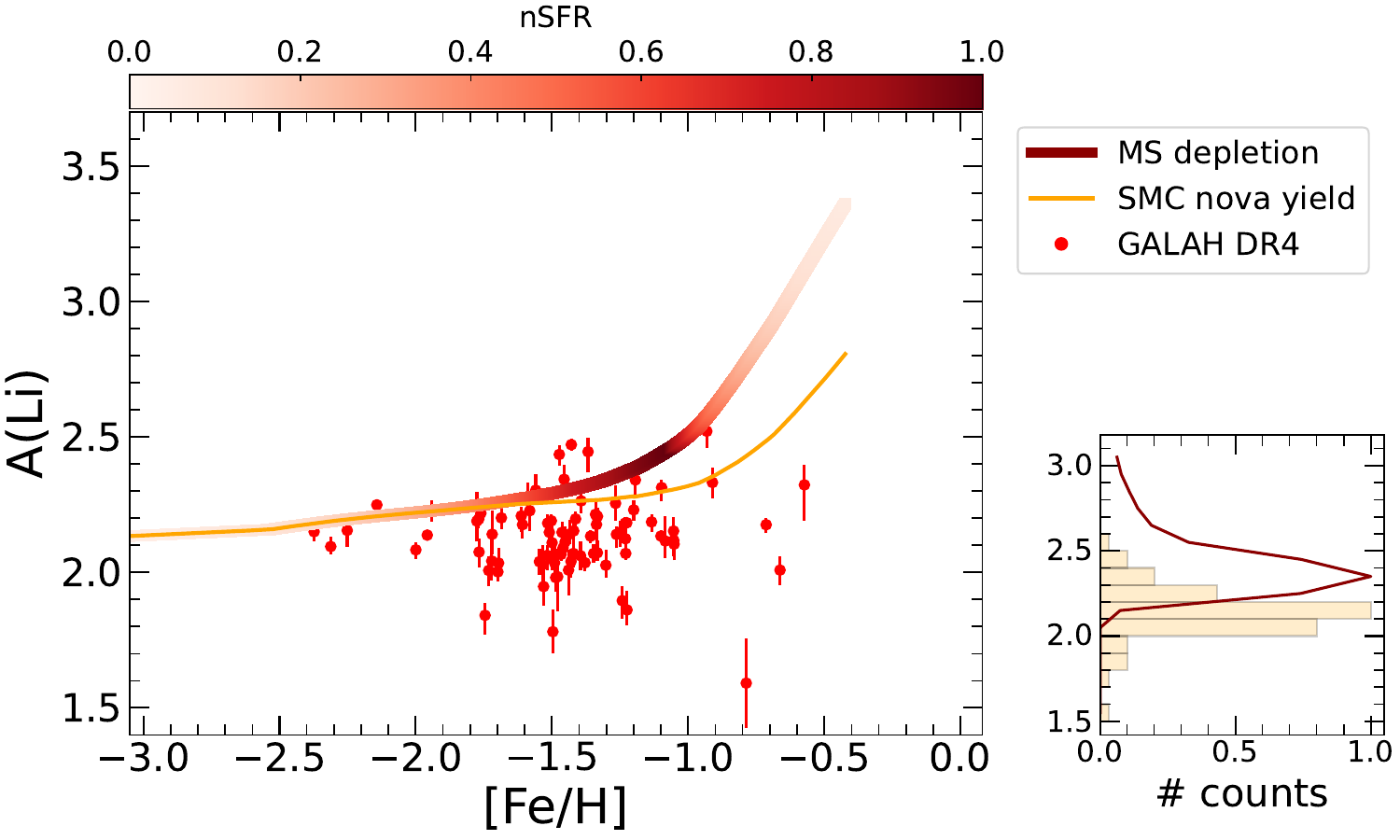}
    \caption{Li evolution in MS stars by using the GALAH DR4 data.}
    \label{DR4_GSE_MS}
\end{figure}

In this appendix we redo our study in this work by using data from the GALAH DR4 catalogue \citep[][]{2025PASA...42...51B}. Overall, the sample size increases to 1842 stars, about 500 stars more than the sample using the GALAH DR3 catalogue. Fig. \ref{DR4_GSE_MDF} shows the obtained MDF, indicating an offset with our model prediction, or rather with the GALAH DR3 catalogue. This may be explained by the quality selection in iron abundances, i.e. we cannot select stars with \texttt{flag\_fe\_h}=0 due to a bug arose in the GALAH DR4. Works in progress to solve this problem and calibrations to further evaluate the new released data have been claimed to be provided in the future \citep[see][]{2025PASA...42...51B}. Regardless, we then make a tentative analysis on the Li evolution. The results are shown in Figs. \ref{DR4_GSE_eRGB} and \ref{DR4_GSE_MS}. For the Li plateaus at low metallicities, we obtain an agreement with our main results in Figs.~\ref{GSE_eRGB} and \ref{GSE_MS}. The nova Li contribution remains ambiguous.

\end{appendix}

\end{document}